\documentclass[preprint,showpacs,preprintnumbers,amsmath,amssymb]{revtex4}

\usepackage{graphicx}

\begin{document}

\title{
Gravitational time delay of light 
for various models of modified gravity 
}
\author{Hideki Asada} 
\email{asada@phys.hirosaki-u.ac.jp}
\affiliation{
Faculty of Science and Technology, Hirosaki University,
Hirosaki 036-8561, Japan} 

\date{\today}

\begin{abstract}
We reexamined the gravitational time delay of light, 
allowing for various models of modified gravity. 
We clarify the dependence of the time delay (and induced frequency shift) 
on modified gravity models and investigate 
how to distinguish those models, 
when light propagates in static spherically symmetric spacetimes. 
Thus experiments by radio signal from spacecrafts 
at very different distances from Sun 
and future space-borne laser interferometric detectors 
could be a probe of modified gravity 
in the solar system. 
\end{abstract}


\maketitle


The nature of dark energy and dark matter has become 
a central issue in modern cosmology. 
Recent observations such as the magnitude-redshift relation 
of type Ia supernovae (SNIa) \cite{SN} 
and 
the cosmic microwave background (CMB) anisotropy by WMAP \cite{WMAP} 
strongly suggest a certain modification, 
in whatever form, in the standard cosmological model. 
We are forced to add a new component into the energy-momentum tensor 
in the Einstein equation 
or modify the theory of general relativity itself 
\cite{Kamionkowski}. 
Indeed, there have been a lot of proposals 
motivated by, for instance, scalar tensor theories, string theories, 
higher dimensional scenarios and quantum gravity 
(For recent reviews of modified gravity models inspired by 
the dark energy observation, e.g.,  \cite{theories}). 
Therefore, it is of great importance to observationally test 
these models. 

The theory of general relativity has passed ``classical'' tests, 
such as the deflection of light, the perihelion shift 
of Mercury and the Shapiro time delay, and also a systematic test 
using the remarkable binary pulsar ``PSR 1913+16'' 
and several binary pulsars now known \cite{Will}. 
In the twentieth century, these tests proved that 
the Einstein's theory is correct with a similar accuracy of $0.1 \%$. 

Since the time delay effect along a light path in the gravitational 
field was first noticed in 1964 by Shapiro \cite{Shapiro}, 
this effect has successfully tested the Einstein's theory 
\cite{Will06}. 
A significant improvement was reported in 2003 from Doppler tracking 
of the Cassini spacecraft on its way to the Saturn, 
with $\gamma-1 = (2.1 \pm 2.3) \times 10^{-5}$  
\cite{Cassini}. 
Here, $\gamma$ is one of parameters in the parameterized 
post-Newtonian (PPN) formulation of gravity 
\cite{Will}. 
The bending and delay of photons by the curvature of 
spacetime produced by any mass are proportional to $\gamma+1$, 
where $\gamma$ is unity in general relativity 
but zero in the Newtonian theory, 
and the quantity $\gamma-1$ is thus considered as a measure 
of a deviation from general relativity. 
The sensitivity in the Cassini experiment approaches 
the level at which, theoretically, deviations $10^{-6} - 10^{-7}$ 
are expected in some cosmological models \cite{DP,DPV}. 
Therefore, it is important to investigate the Shapiro time delay 
with such a high accuracy. 

In addition to the above theoretical motivation, 
there are advances in technologies concerning 
the high precision measurement of time and frequency 
such as optical lattice clocks \cite{Katori} and 
attoseconds ($10^{-18}$ s) laser technologies \cite{Sansone}. 
ASTROD project with three spacecrafts aims at measuring 
$\gamma$ at the level of $10^{-9}$ \cite{Ni}. 

The purpose of this paper is to clarify the dependence of 
the time delay (and induced frequency shift) 
on modified gravity models and investigate 
how to distinguish those models by using the Shapiro time delay. 
An important point in this paper is that 
we allow for various modified gravity theories 
beyond the scope of the PPN formulation. 
Introducing a new energy or length scale 
(e.g. extra dimension scale) 
may make changes in functional forms of 
the gravitational field. 
Thus it is worthwhile to investigate how to probe 
such a modified functional form, by using the light propagation 
in the solar system. 
Throughout this paper, we take the units of $G=c=1$.

In this paper, we assume that the electromagnetic fields 
propagate in four-dimensional spacetimes 
(even if the whole spacetime is higher dimensional). 
Thus photon paths follow null geodesics 
(as the geometrical optics approximation of Maxwell equation). 

We shall consider a static spherically symmetric spacetime, 
in which light propagates, 
expressed as  
\begin{equation}
ds^2=-A(r) dt^2 + B(r) dr^2 + r^2 d\Omega^2 , 
\end{equation}
where $r$ and $d\Omega^2$ denote the circumference radius 
and the metric of the unit 2-sphere, respectively. 
The functions $A(r)$ and $B(r)$ depend on gravity theories. 

The time lapse along a photon path is obtained as 
\begin{equation}
t(r,r_0)
=\int^r_{r_0} \frac{dr}{b} \sqrt{\frac{B(r)}{A(r)}} 
\frac{1}{\sqrt{\frac{A(r_0)}{r_0^2}-\frac{A(r)}{r^2}}} ,
\end{equation}
where $b$ and $r_0$ denote the impact parameter and the closest point, 
respectively. Their relation is $b^2=r_0^2/A(r_0)$. 

According to a concordance between solar-system experiments and 
the theory of general relativity, we can assume that the spacetime 
is expressed as the Schwarzschild metric (rigorously speaking, 
its weak field approximation)  
with a small perturbation induced by modified gravity. 
For practical calculations, we keep only the leading term 
at a few AU in the corrections.  
Namely, $A(r)$ and $B(r)$ are approximated as 
\begin{eqnarray}
A(r)&\approx&1-\frac{2M}{r}+A_m r^m ,
\label{A}
\\
B(r)&\approx&1+\frac{2M}{r}+B_n r^n , 
\label{B}
\end{eqnarray}
where $M$ denotes the mass of the central body.  
Here, $A_m$, $B_n$, $m$ and $n$ rely on a theory which 
we wish to test.  
For simplicity, we assume $m = n > 0$, which corresponds to 
a wide class of theories of gravity. 

Examples of modified gravity theories are as follows. 
(1) $n=1/2$, $A_n = -2 B_n = \pm 2 \sqrt{M/r_c^2}$ for DGP model 
with $r_c$ is the extra scale within which gravity becomes 
five dimensional 
\cite{DGP}. 
(2) $n = 3/2$, $A_n = (2/3) m_g^2 \sqrt{2 M/13}$ and 
$B_n= - m_g^2 \sqrt{2 M/13} $ with graviton mass $m_g$ 
for one of massive gravity models \cite{MG1,MG2}. 
(3) $n=2$, $A_n = -B_n = - \Lambda / 3$ 
for the Schwarzschild-de Sitter spacetime, 
that is, general relativity with the cosmological constant 
$\Lambda$ as a possible candidate for the dark energy, 
though this is not a manifest modification of gravity. 
The solar system experiments are not sensitive to 
this model with $\Lambda \sim 10^{-52} m^{-2}$ \cite{KKL}.  
Here, it should be noted that the examples (1) and (2) 
give conformally flat spacetimes (in the weak field approximation) 
and their conformal factors generate the gravitational time delay 
(and induced frequency shift), though the null geodesic 
in any conformally flat spacetime 
is mapped into that in the Minkowski one. 
 
The Cassini experiment has put the tightest constraint 
on the solar gravity, especially near the solar surface 
with the accuracy of $10^{-5}$ \cite{Cassini}. 
This implies that deviations in $A(r_{\odot})$ and 
$B(r_{\odot})$ must be less than 
$10^{-5} \times 2 M_{\odot}/r_{\odot} \sim 10^{-10}$, 
that is, 
$|A_m r_{\odot}^m|$, $|B_n r_{\odot}^n| < 10^{-10}$.  

We consider the round-trip time between pulse transmission 
and echo reception, denoted by $\Delta T$. 
The pulse is emitted from Earth at $r_E$, 
and reflected at $r_R$. 

Up to the linear order in $M$, $A_n$ and $B_n$, 
$\Delta T$ is expressed as 
\begin{eqnarray}
\Delta T &=& 
2 \left( \sqrt{r_E^2-r_0^2}+\sqrt{r_R^2-r_0^2} \right)
\nonumber\\
&&+2 M \left( 2\ln{\frac{r_E+\sqrt{r_E^2-r_0^2}}{r_0}}
+2\ln{\frac{r_R+\sqrt{r_R^2-r_0^2}}{r_0}}
+\sqrt{\frac{r_E-r_0}{r_E+r_0}}
+\sqrt{\frac{r_R-r_0}{r_R+r_0}}
 \right)
\nonumber\\
&&+\delta t . 
\label{Deltat}
\end{eqnarray}
The extra time delay induced by a correction to general relativity 
is expressed as 
\begin{eqnarray}
\delta t &=& 
r_0^{n+1} 
\left( \int_{1}^{R_E}+\int_{1}^{R_R} \right) dR 
\nonumber\\
&&
\times \left(
-A_n\frac{R^{n+3}-2 R^{n+1}+R}{(R^2-1)^{3/2}}
+B_n\frac{R^{n+1}}{\sqrt{R^2-1}}
\right) , 
\label{deltat}
\end{eqnarray}
where we define nondimensional radial coordinates as 
$R \equiv r/r_0$, 
$R_E \equiv r_E/r_0$ 
and 
$R_R \equiv r_R/r_0$. 
For a radar tracking of a spacecraft such as Cassini, 
$r_E$ and $r_R$ are of the order of 1 AU ($\sim 10^{8}$ km), 
and $r_0$ is several times of the solar radius 
($R_{\odot} \sim 10^5$ km). 
Equation ($\ref{deltat}$) can be rewritten by using 
special functions, though it seems less informative. 
Therefore, we take expansions of Eq. ($\ref{deltat}$) 
in $r_0$ because of $r_E, r_R \gg r_0$. 
For $n \neq 1$, we obtain 
\begin{eqnarray}
\delta t &=& 
\frac{B_n-A_n}{n+1}
\left( r_E^{n+1}+r_R^{n+1} \right) 
+\frac{B_n+A_n}{2 (n-1)} 
\left( r_E^{n-1}+r_R^{n-1}-2 r_0^{n-1} \right) r_0^2 
+O(r_0^4) ,    
\label{deltat2}
\end{eqnarray} 
whereas the second term of R.H.S. becomes  
$(B_n+A_n) \ln(r_E r_R/r_0^2) r_0^2/2$ 
for $n=1$.

It is convenient to use the relative change in the frequency, 
which is caused by the gravitational time delay \cite{Bertotti}, 
because the Doppler shift due to the receiver's motion has no effect 
owing to the cancellation at both the receipt and emission of 
radio signal \cite{Bertotti}. 
This frequency shift is defined as 
$y=- d(\Delta T)/dt$. 
Indeed, the frequency shift was used by the Cassini experiment. 
For a case of $b \ll r_E, r_R$, which is valid for the Cassini experiment,  
the general relativistic contribution is expressed as \cite{Will} 
\begin{eqnarray}
y_{GR} = 4 \frac{M}{b} \frac{db}{dt} . 
\label{yGR}
\end{eqnarray}

We pay attention to the extra contribution 
due to modified gravity. 
For $n \neq 1$, the extra frequency shift becomes 
\begin{eqnarray}
\delta y = - \frac{A_n+B_n}{n-1} 
\{ r_E^{n-1}+r_R^{n-1}-(n+1)r_0^{n-1} \} 
b \frac{db}{dt} ,   
\label{deltay}
\end{eqnarray}
while we obtain 
$\delta y = -(A_n+B_n) [ \ln(r_E r_R/r_0^2) -1] b db/dt$ 
for $n=1$. 
Here we used $dr_E/dt$, $dr_R/dt \ll dr_0/dt$ $(\sim db/dt)$ 
near the solar conjunction ($b \ll r_E, r_R$).  
The total frequency shift $y$ is the sum, $y_{GR} + \delta y$. 
The impact parameter of light path changes with time, 
because of the motion of the emitter and receiver 
with respect to Sun. 
For simplicity, we assume that they move at constant velocity 
during short-time observations. 
The impact parameter changes as 
$b(t)=\sqrt{b_0^2+v^2 t^2}$, 
where $b_0$ denotes the minimum of the impact parameter 
near the solar conjunction at $t=0$, 
and $v$ is the velocity component perpendicular to 
the line of sight.


Here, we make an order-of-magnitude estimate of the frequency shift. 
First, we obtain 
$y_{GR} \sim 10^{-9} (M/M_{\odot}) (r_{\odot}/b) (\dot{b}/v_E)$, 
where 
the dot denotes the time derivative, 
and $v_E$ is the orbital velocity of Earth ($\sim$ 30 km/s). 
The Cassini experiment 
reported $y$ at the level of $10^{-14}$ by careful processing 
of the frequency fluctuations 
largely due to the solar corona and the Earth's troposphere 
\cite{Cassini}. 
Multi-band measurements are preferred in order to 
avoid the astrophysical effect of the corona and 
interplanetary plasma on the delay, 
which is proportional to the square inverse of the frequency. 

For a receiver at $r_R > r_E$,  
the extra frequency shift is 
\begin{eqnarray}
\delta y &\sim& (A_n+B_n) r_R^n 
\frac{b}{r_R} \frac{db}{dt}
\nonumber\\
&\sim& 
10^{-17} \left(\frac{10 \mbox{AU}}{r_{\odot}}\right)^n 
\left(\frac{(A_n+B_n) r_{\odot}^n}{10^{-10}}\right)
\left(\frac{r_R}{10 \mbox{AU}}\right)^{n-1} 
\left(\frac{b}{r_{\odot}}\right) 
\left(\frac{db/dt}{v_E}\right) ,  
\label{order}
\end{eqnarray}
where $10 \mbox{AU} / r_{\odot} \sim 2 \times 10^3$.  
The larger the index of $n$, the longer the delay $\delta y$. 

\begin{figure}
\includegraphics[width=14cm]{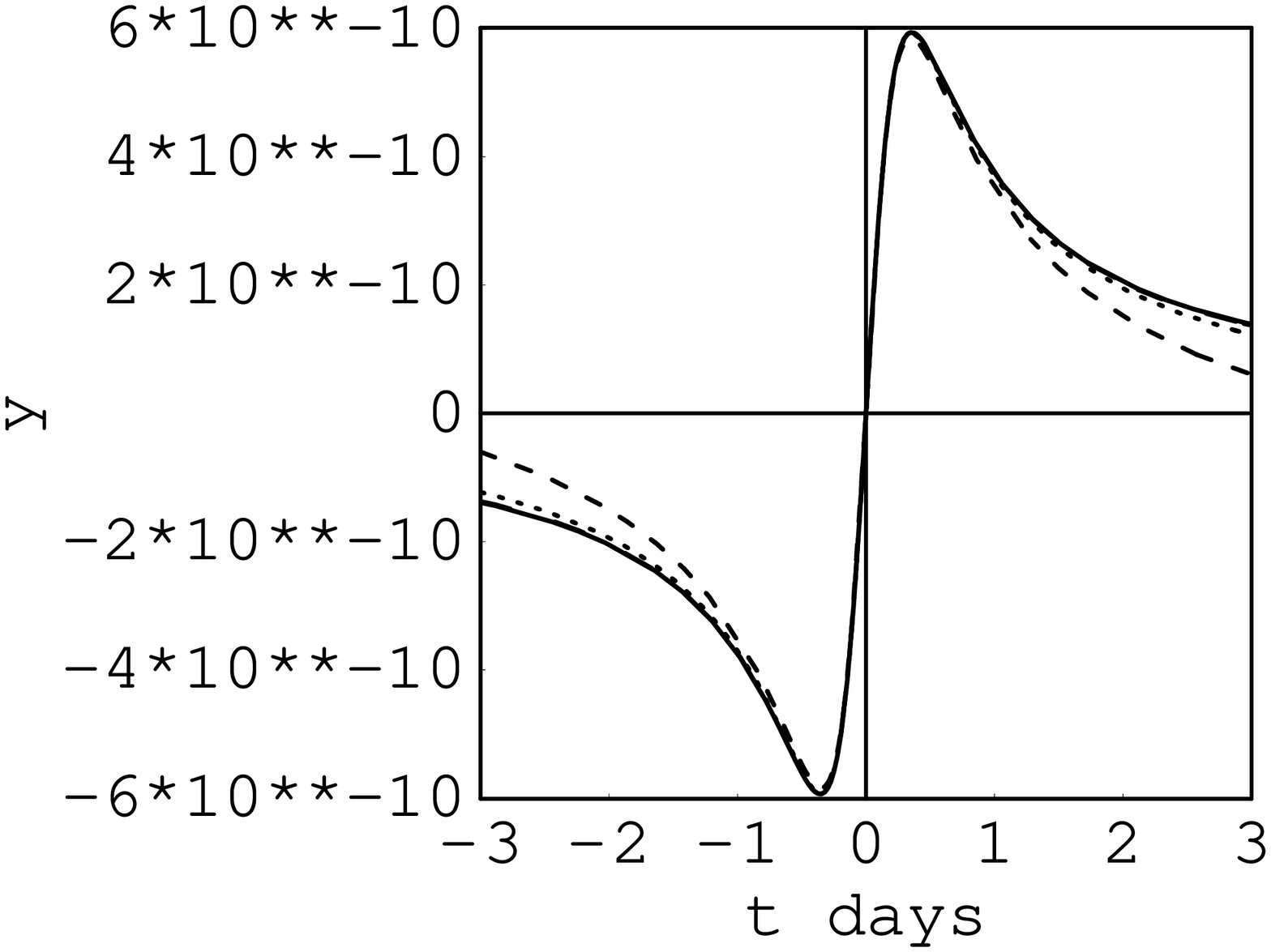}
\caption{
Dependence of the frequency shift on the distance $r_R$ and 
the index $n$. 
The long dashed,  short dashed and dotted curves denote 
the frequency shift for $(n, r_R)=(3/2, 10 \mbox{AU})$, 
$(n, r_R)=(2, 10 \mbox{AU})$, 
$(n, r_R)=(2, 1 \mbox{AU})$, respectively. 
The long dashed curve for $n=3/2$ and $r_R = 10$ AU 
is overlapped with the solid curve denoting 
the general relativistic case. 
Here, we assume $(A_n+B_n) r_{\odot}^n = 3 \times 10^{-11}$. 
}
\label{figure-shift2}
\end{figure}

Figure $\ref{figure-shift2}$ shows that an extra distortion 
due to $\delta y$ would appear especially 
in the tail parts of $y-t$ curves. 
According to the fact that no deviation from general relativity 
has been reported by the Cassini experiment \cite{Cassini},  
we can put a constraint as $\delta y < 10^{-14}$ 
at $r_R = 8.43$AU. 
On the other hand, Eq. $(\ref{order})$ gives 
$\delta y \sim 10^{-11}$ for 
$n=2$ and $(A_n+B_n) r_{\odot}^n = O(10^{-10})$, 
for instance, which are thus rejected. 
One can distinguish modified gravity
models, which are characterized by various values of 
$n$, $A_n$, $B_n$, from observations using receivers 
at very different distances from Sun, 
as shown by Fig. $\ref{figure-shift2}$. 

\begin{figure}
\includegraphics[width=14cm]{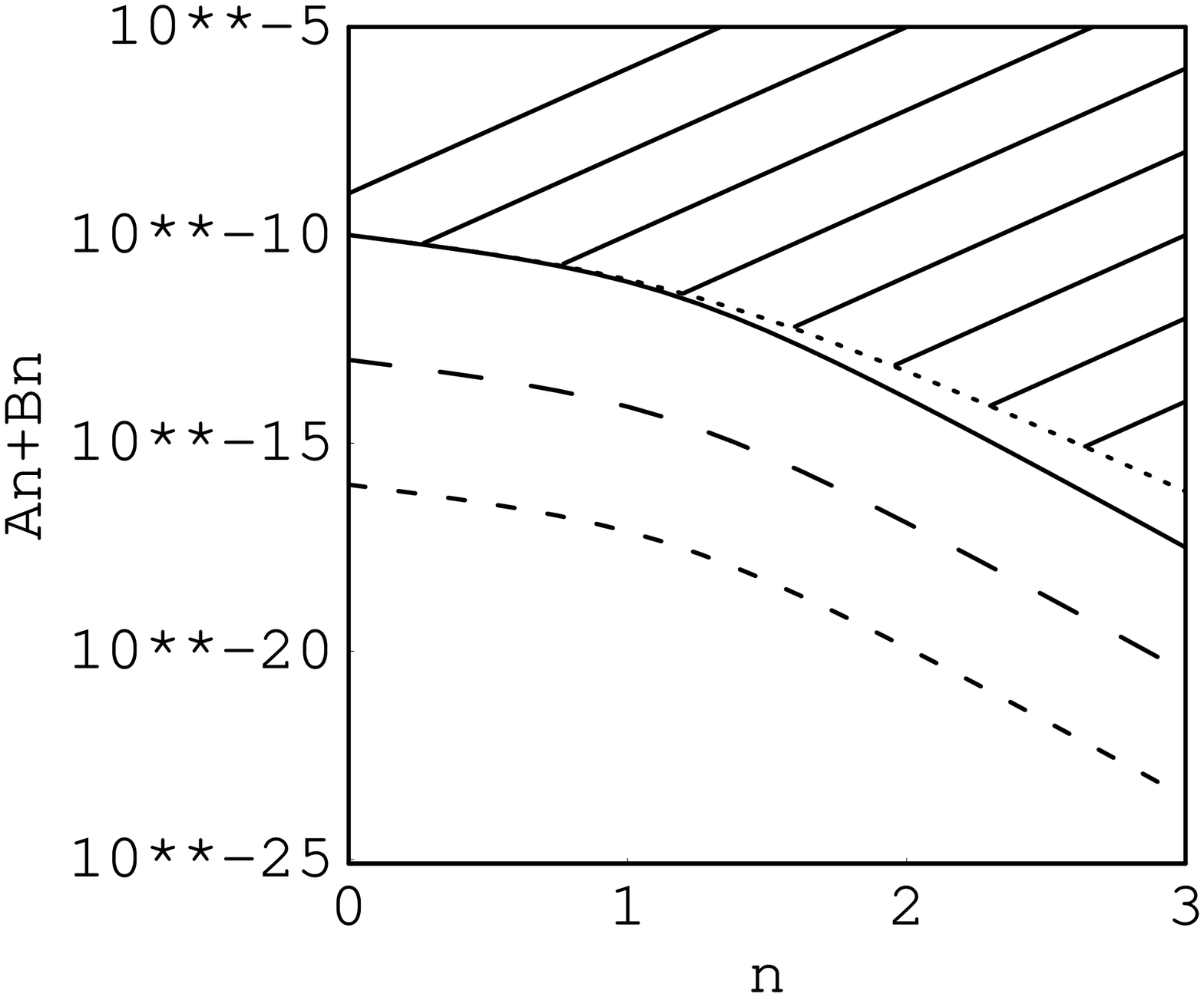}
\caption{
Contours of $\delta y$ on the $n$ - $|A_n+B_n| r_{\odot}^n$ plane. 
The solid, long-dashed and short-dashed curves correspond to 
$\delta y = 10^{-14}, 10^{-17}, 10^{-20}$, respectively, 
where we assume $r_E = 1$ AU, $r_R = 40$ AU, $b \sim r_{\odot}$ 
and $db/dt \sim v_E$. 
The limit due to the current technology is $\delta y \sim 10^{-17}$. 
The shaded region above the dotted curve 
($\delta y = 10^{-14}$ for $r_R=8.43$ AU) 
is excluded, because no devitation up to $O(10^{-14})$ 
has been detected by the Cassini experiment \cite{Cassini}. 
}
\label{figure-constraint}
\end{figure}

Figure $\ref{figure-constraint}$ shows the dependence 
of $\delta y$ on $n$ and $A_n+B_n$. 
Hence, one can put a constraint on $n$ and $A_n+B_n$ from 
$\delta y$ observed.

Equation ($\ref{yGR}$) shows that the frequency shift 
depends only on the impact parameter $b$ but not the locations 
of the emitter and receiver. 
Strictly speaking, $y_{GR}$ still has 
weak dependence on $r_E$ and $r_R$ as shown by Eq. ($\ref{Deltat}$). 
On the other hand, $\delta y$ depends strongly on $r_E$ and $r_R$. 
The dependence of $y_{GR}$ and $\delta y$ on $r_E$ and $r_R$ 
plays a crucial role in constraining (or detecting) 
a correction to general relativity in the solar system. 

Let us imagine that time delays (or induced frequency shifts) 
are measured along two light trajectories, whose impact parameters 
are denoted as $b_1$ and $b_2$, respectively.  
Then, we make a comparison of the two time delays.  
If they are in good agreement after taking account of 
a difference in the impact parameters, 
general relativity can be verified again. 
Otherwise, a certain modification could be required 
for the solar gravitational field. 
At this stage, however, one can say nothing about functional forms 
of the correction because 
the parameters of both $n$ and $A_n+B_n$, which we wish to determine, 
enter the frequency shift.  

In order to break this degeneracy, therefore, 
we consider three light paths, for which 
the impact parameters of the photon paths are almost 
the same (several times of the solar radius) for convenience sake. 
The locations of the receivers are denoted as 
$r_{R1}$, $r_{R2}$ and $r_{R3}$, where 
the subscripts from 1 to 3 denote each light path. 
We assume that $r_E$ is constant in time for simplicity. 
It is a straightforward task to take account of the eccentricity 
of the Earth orbit and a difference between the impact parameters. 

We make use of a difference such as  
$y_2 - y_1$ and $y_3 - y_1$, in order to 
cancel out general relativistic parts.   
We find 
\begin{equation}
y_2-y_1 = \frac{A_n+B_n}{n-1} 
(r_{R1}^{n-1}-r_{R2}^{n-1})b\frac{db}{dt} .  
\label{difference-y}
\end{equation}
It should be noted that $y_2-y_1$ is proportional to $A_n+B_n$. 
Hence, the following ratio depends only on $n$ as 
\begin{equation}
\frac{y_3-y_1}{y_2-y_1}
=\frac{r_{R1}^{n-1}-r_{R3}^{n-1}}{r_{R1}^{n-1}-r_{R2}^{n-1}} . 
\end{equation}
Thereby, one can determine the index $n$. 
Next, one obtains $A_n+B_n$  
by substituting the determined $n$ into Eq. ($\ref{difference-y}$).

In summary, 
we have clarified the dependence of the gravitational time delay 
on modified gravity models. 
For neighboring light rays, Eq. ($\ref{deltat2}$) gives 
almost the same value so that one can hardly distinguish 
models of gravity. 
This implies that we should prepare receivers 
at very different distances from Sun. 
Our result could be used for exterior planets explorers 
such as New Horizons, which were launched in 2006 
and their primary target is Pluto and its moon, Charon 
at distance from Sun $\sim$ 40 AU \cite{NH}.   
In future practical data analyses, however, it would be safer 
to use the original integral form as Eq. ($\ref{deltat}$), 
because Eq. ($\ref{deltat2}$) is an approximate expression. 

Furthermore, $b$ becomes the same order of $r_E$, $r_R$ 
for future space-borne laser interferometric detectors 
such as LISA \cite{LISA}, DECIGO \cite{DECIGO} and 
especially ASTROD \cite{Ni}. 
These detectors are in motion in our solar system. 
Namely, $r_R$ and $b$ change with time. 
Therefore, the sophisticated experiments by space-borne laser 
interferometric detectors, which are originally designed to detect 
time-dependent part of gravity, {\it i.e.} gravitational waves, 
could probe also a time-independent 
part of gravity at the relative level of 
$\Delta y \sim \Delta\nu/\nu \sim \Delta L/L < 10^{-20}$.  
It would be important to make a feasibility estimate for 
these detectors. 
Clearly, a stronger test can be done by not a single experiment 
but combining several ones.   
In addition, more precise measurements of the Shapiro time delay with 
binary pulsars may put a constraint on the modifications discussed 
in this paper, especially in the strong self-gravitating regime. 
Further investigations along these lines will be done in the future. 

\acknowledgements
The author would like to thank S. Kawamura, N. Mio, M. Sasaki, 
M. Shibata and T. Tanaka for useful conversations. 
This work was supported by a Japanese Grant-in-Aid 
for Scientific Research from the Ministry of Education, 
No. 19035002.

\begin{thebibliography}{99}
\bibitem{SN}
S.~Perlmutter {\it et al.},  
  Astrophys.\ J.\  {\bf 517}, 565 (1999); 
  A.~G.~Riess {\it et al.},  
  Astrophys.\ J.\  {\bf 607}, 665 (2004). 
\bibitem{WMAP}
D.~N.~Spergel {\it et al.},  
  Astrophys.\ J.\ Suppl.\  {\bf 148}, 175 (2003). 
\bibitem{Kamionkowski}
M.~Kamionkowski, 
arXiv:0706.2986. 
\bibitem{theories}
S.~Nojiri and S. D.~Odintsov, 
Int. J. Geom. Meth. Mod. Phys. {\bf 4}, 115 (2007);  
  E.~J.~Copeland, M.~Sami and S.~Tsujikawa,
  Int.\ J.\ Mod.\ Phys.\  D {\bf 15}, 1753 (2006); 
V.~Sahni and A.~Starobinsky,
  Int.\ J.\ Mod.\ Phys.\  D {\bf 15}, 2105 (2006). 
\bibitem{Will}
C. M. Will, {\it Theory and experiment in gravitational physics} 
(Cambridge Univ. Press, Cambridge, 1993). 
\bibitem{Shapiro}I. I. Shapiro, Phys. Rev. Lett. {\bf 13}, 789 (1964).
\bibitem{Will06}
C. M. Will, Living Rev. Relativity 9, 3 (2006), 
http://relativity.livingreviews.org/Articles/lrr-2006-3.
\bibitem{Cassini}B. Bertotti, L. Iess and P. Tortora, 
Nature, {\bf 425}, 374 (2003). 
\bibitem{DP}
T.~Damour and A.~M.~Polyakov,
  Nucl.\ Phys.\  B {\bf 423}, 532 (1994). 
\bibitem{DPV}
T.~Damour, F.~Piazza and G.~Veneziano,
  Phys.\ Rev.\  D {\bf 66}, 046007 (2002). 
\bibitem{Katori}
M. Takamoto, F.~L. Hong, R. Higashi, H. Katori,
Nature, {\bf 435}, 321 (2005). 
\bibitem{Sansone}
G. Sansone et al., Science, {\bf 314}, 443 (2006). 
\bibitem{Ni}
  W.~T.~Ni {\it et al.},
  J.\ Phys. 
{\bf 32}, 154 (2006).
\bibitem{DGP}
G.~R.~Dvali, G.~Gabadadze and M.~Porrati,
  Phys.\ Lett.\  B {\bf 485}, 208 (2000). 
\bibitem{MG1}
A.~I.~Vainshtein,
  Phys.\ Lett.\  B {\bf 39}, 393 (1972). 
\bibitem{MG2} 
T.~Damour, I.~I.~Kogan and A.~Papazoglou,
  Phys.\ Rev.\  D {\bf 67}, 064009 (2003). 
\bibitem{KKL}
V.~Kagramanova, J.~Kunz and C.~Lammerzahl,
  Phys.\ Lett.\  B {\bf 634}, 465 (2006). 
\bibitem{Bertotti}
B. Bertotti and G. Giampieri, 
Class. Quant. Grav. {\bf 9}, 777 (1992); 
L. Iess, G. Giampieri, J. D. Anderson and B. Bertotti, 
Class. Quant. Grav. {\bf 16}, 1487 (1999).
\bibitem{NH}http://www.nasa.gov/mission\underline{\;}
pages/newhorizons/main/index.html
\bibitem{LISA}http://lisa.nasa.gov/
\bibitem{DECIGO}
  N.~Seto, S.~Kawamura and T.~Nakamura,
  Phys.\ Rev.\ Lett.\  {\bf 87}, 221103 (2001); 
S.~Kawamura {\it et al.},
  Class.\ Quant.\ Grav.\  {\bf 23}, S125 (2006).
\end{thebibliography}
\end{document}